\documentclass[a4paper]{article}
\usepackage[dvips]{graphics}
\usepackage{mathptm}  
\usepackage{times}
\begin{document}

 \title{ Generalized Optimal Current Patterns and Electrical Safety in EIT}
\author{William R.B. Lionheart\thanks{%
~Department of Mathematics, UMIST, UK
} \and
Jari Kaipio  \thanks{%
~Department of Applied Physics, University of Kuopio,
Finland} \and
Christopher N. McLeod\thanks{%
~School of Engineering, Oxford Brookes
University,UK } }
\maketitle

\begin{abstract}
  There are a number of constraints which limit the current and voltages which
  can be applied on a multiple drive electrical imaging system.  One
obvious  constraint is to limit the maximum Ohmic power dissipated in
the body.  Current patterns optimising distinguishability with
respect to this constraint are singular functions of the difference
of transconductance  matrices with respect to the power norm. 
(the optimal currents
of Isaacson).  If one constrains the  total current ($L^1$ norm)  the
optimal patterns are pair drives.  On the other  hand if one
constrains the maximum current on each drive electrode (an
$L^\infty$ norm), the optimal patterns have each drive channel set
to the  maximum source or sink current value.
In this paper we  consider appropriate safety constraints and discuss how to find the
optimal current patterns with those  constraints.
\end{abstract}

\section{Introduction}
The problem of optimizing the drive patterns in EIT was first
considered by Seagar~\cite{Seagar:Thesis} who calculated the optimal
placing of a pair of point drive electrodes on a disk to maximize  the
voltage differences between the measurement of a homogeneous background
and an offset circular anomaly. Isaacson~\cite{Isaacs}, and Gisser, Isaacson and Newell~\cite{GIN}
argued that one should maximize the $L^2$ norm of the voltage
difference between the measured and calculated voltages constraining
the $L^2$ norm of the current patterns in a multiple drive
system. Later~\cite{GIN2} they used a constraint on the maximum dissipated power
in the test object. Ey\"{o}bo\v{g}lu and Pilkington~\cite{EP} argued that medical
safety legislation demanded that one restrict  the maximum total current entering the body, and if
this constraint was used the distinguishability is maximized by pair
drives. 
Cheney and Isaacson~\cite{CI} study a concentric anomaly in a disk,
using the 'gap' model for electrodes. They compare trigonometric,
Walsh, and opposite and adjacent pair drives for this case giving the
dissipated power as well as the $L^2$ and power distinguishabilies. 
K\"{o}ksal and Ey\"{o}bo\v{g}lu~\cite{KE} investigate the concentric
and offset anomaly in a disk using continuum currents.  

Yet another approach~\cite{Breckon} is to  find a
current pattern maximizing the voltage difference for a single
differential voltage measurement.

\section{Medical Electrical Safety Regulations}

We will review the current safety regulations  here, but notice that
they were not designed with multiple drive EIT systems in mind and we
hope to stimulate a debate about what would be appropriate safety standards.

For the purposes of this discussion the equipment current (``Earth Leakage 
Current" and ``Enclosure Leakage Current") will be ignored as the emphasis is on 
the patient currents. These will be assessed with the assumption that the 
equipment has been designed such that the applied parts, that is the electronic 
circuits and connections which are attached to the patient for the delivery of 
current and the measurement of voltage, are fully isolated from the protective 
earth (at least $50M\Omega$).
 
IEC601 and the equivalent BS5724 specify a safe limit of 100 $\mu$A   for 
current flow to protective earth (``Patient Leakage Current") through electrodes 
attached to the skin surface (Type BF) of patients under normal conditions. 
This is designed to ensure that the equipment will not put the patient at risk 
even when malfunctioning. The standards also specify that the equipment should 
allow a return path to protective earth for less than 5 mA if some other 
equipment attached to the patient malfunctions and applies full mains voltage 
to the patient. Lower limits of 10 $\mu$A (normal) and 50 $\mu$A (mains 
applied to the patient) are set for internal connections, particularly to the 
heart (Type CF), but that is not at present an issue for EIT researchers. 
 
The currents used in EIT flow between electrodes and are described in the 
standards as ``Patient Auxiliary Currents" (PAC). The limit for any PAC is a 
function of frequency, 100 microamps from 0.1Hz to 1 kHz; then $100f$ $\mu$A
 from 1 kHz  to 100 kHz where $f$ is the frequency in kHz; then 10 mA 
above 100 kHz. The testing conditions for PAC cover 4 configurations; the worst 
case of each should be examined.

1. Normal conditions. The design of single or multiple current source 
tomographs should ensure that each current source is unable to apply more than 
the maximum values given.
 
2. The PAC should be measured between any single connection and all the other 
connections tied together. 
a) if the tomograph uses a single current source then the situation is similar 
to normal conditions (above)
b) if the tomograph uses multiple current sources then as far as the patient is 
concerned the situation is the same as normal conditions. The design of the 
sources should be such that they will not be harmed by this test.
 
3. The PAC should be measured when one or more electrodes are disconnected from 
the patient. This raises issues for multiple-source tomographs :
a) if an isolated-earth electrode is used then the current in it will be the 
sum of the currents which should have flowed in the disconnected electrodes; 
they could all be of the same polarity. The isolated-earth electrode should 
therefore include an over-current sensing circuit which will turn down/off all 
the current sources.
b) If no isolated-earth electrode is used then the situation is similar to 
normal conditions.
 
4. The PAC should be measured when the disconnected electrodes are connected to 
protective earth. This introduces no new constraints given the tomograph is 
fully isolated.

\section{Constrained Optimization}
Let $V=(V_1,\dots,V_K)^T$ be the vector of potentials measured on electrodes when a
pattern of currents $I=(I_1,\dots,I_K)^T$ is applied. These are related
linearly by $R$ the
transfer impedance matrix: $V=RI$. For
simplicity we will assume the same system of electrodes is used for
current injection and voltage measurement. We will also assume that
the conductivity is real and the currents in-phase to simplify the
exposition. A model of the body is used
with our present best estimate for the conductivity and from this we
calculate 
voltages $V_\mathrm{c}$ for the same current pattern. Our aim is to
maximize the distinguishability $\|V- V_\mathrm{c}\|_2=\|(R- R_\mathrm{c})I\|_2$. The use of
the $L^2$ norm here corresponds to the assumption that the noise on
each measurement channel is independent and identically
distributed. If there were no constraints on the currents the
distinguishability would be unbounded.

The simplest idea~\cite{GIN} is  to maximize $\|(R-R_\mathrm{c})I\|_2$
subject to $\| I \|_2 \le M$ for some fixed value of $M$. The solution of
this problem that $I$ is the eigenvector of $R-R_\mathrm{c}$
corresponding to the largest (in absolute value) eigenvalue. One
problem is that the 2-norm of the current has no particular
physical meaning. In a later paper~\cite{GIN2} it was proposed that
the dissipated power be constrained, that is $I\cdot V= I^T R I$. The
optimal current is the  eigenvector of
$(R-R_\mathrm{c})R^{-1/2}$. (The inverse implied in the expression
$R^{-1/2}$ has to be understood in the generalized sense, that is one
projects on to the space orthogonal to $(1,\dots,1)^T$ and then
calculates the matrix exponent $1/2$.)  In practical situations in medical EIT
the total dissipated power is unlikely to be an active constraint,
although local heating effects in areas of high current density may be
an issue. Even in industrial applications of EIT, the limitations of
voltages and currents handled by normal electronic devices mean that
one is unlikely to see total power as a constraint. One exception
might be in EIT applied to very small objects.

As we have seen a reasonable interpretation of the safety regulations
is to limit the current on each electrode to some safe level
$I_\mathrm{max}$. We will refer to this as an $L^\infty$ constraint. 
 This corresponds to a convex system of linear
constraints $-I_\mathrm{max} \le I_k \le I_\mathrm{max}$. When we
maximize the square of the distinguishabilty, which is a positive
definite quadratic function of $I$, with respect to this set of
constraints it can be seen  that the maximum must be a vertex of the
convex polytope $\{ I:\max_k\{ |I_k|\}=I_\mathrm{max},\sum_k I_k=0\}$.
For example, for an even number $2n$ of electrodes the ${}^{2n}C_n$ vertices are the
currents with each $I_k = \pm I_\mathrm{max}$, and an equal number
with each sign. For the circularly symmetric case these are the  Walsh
patterns referred to in ~\cite{CI}.

If one wanted to be safe under the multiple fault condition that all
the electrodes driving a current with the same sign became
disconnected, and the safety mechanism on the isolated-earth failed,
one would employ the $L^1$ constraint $\sum_k |I_k| \le 2
I_\mathrm{max}$. Again this gives a convex feasible set. In this case
a polyhedron with vertices $I$ such that all but two $I_k$ are zero,
and those two are $I_\mathrm{max}$ and $-I_\mathrm{max}$. These are
the pair drives as considered by Seagar. Pair drives were also
considered by~\cite{CI},\cite{EP},\cite{KE} for single circular
anomalies. Notice that $L^1$ optimal currents will be pair drives for
any two- or three-dimensional geometry and any conductivity distribution.

Another constraint which may be important in practice is that the
current sources are only able to deliver a certain maximum voltage 
$V_\mathrm{max}$
close to their power supply voltage. If the EIT system is connected to
a body with transfer impedance within its design specification then
the constraints $-V_\mathrm{max}\le V_k \le V_\mathrm{max}$ will not be
active. If they do become active then the additional linear
constraints in $I$ space $-V_\mathrm{max}\le R^{-1}I \le V_\mathrm{max}$
(here $R^{-1}$ is to be interpreted as the generalized inverse), will still
result in a convex feasible region.

When any of the linear constraints are combined with quadratic
constraints such as maximum power dissipation the feasible set of
currents is still convex but its surface is no longer a polytope.

\section{Numerical Results}

Although we can easily find  the vertices of the feasible region there
are too many for it to be wise to search exhaustively for a maximum of
the distinguishability. For $32$ electrodes for example there are
${}^{32}C_{16} >  6\times 10^{8}$. Instead we use a discrete steepest ascent search
method of the feasible vertices. That is from a given vertex we
calculate the objective function for all vertices obtained by changing
a pair of signs, and move to whichever vertex has the greatest
value of the objective function. For comparison we also calculated the
$L^2$ optimal currents, the optimal currents for the power constraint,
and the optimal pair drive ($L^1$ optimal).

\begin{figure}
 \begin{center}
   \includegraphics{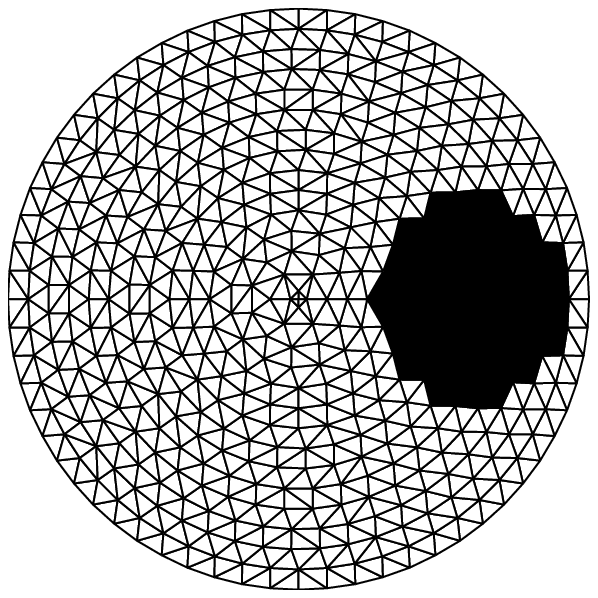}\includegraphics{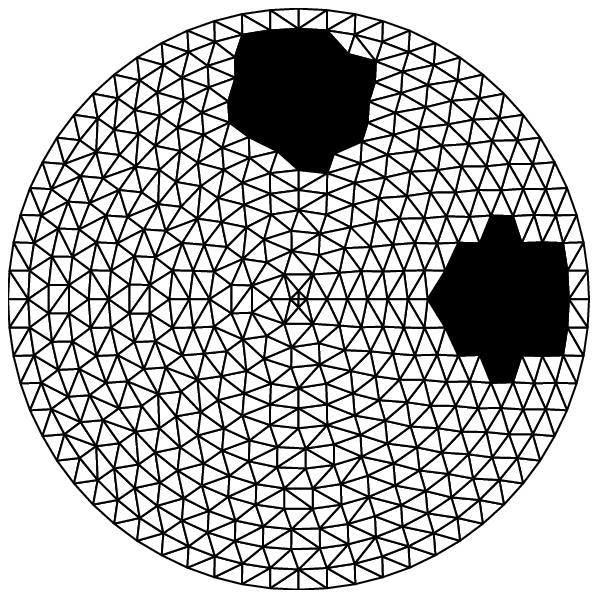}
 \end{center}
 \caption{\label{meshes}Mesh and conductivity anomalies.}
\end{figure}

We used a circular disk for the forward problem, and the
EIDORS Matlab toolbox~\cite{Marko} for mesh generation and forward solution. The
mesh and conductivity targets can be seen in Figure~\ref{meshes}. Our
results are interesting in that for the cases we have studied so far  
the $L^\infty$ optimal currents have  only two sign changes. The
distinguishabilies given in Table~\ref{results} should be read with
caution, as it is  somewhat unfair to compare for example power
constrained with $L^\infty$ patterns. They are designed to optimise
different criteria. However the contrast between pair drive and
$L^\infty$ is worth noting  as the majority of existing EIT systems
can only drive pairs of electrodes.

The greatest current densities occur at the contact points between the
electrode 
boundaries and skin.
At each electrode this current density is determined mainly by the total electrode current, the contact
impedance and the skin conductivity just below the electrode. These factors dominate the current
density near the electrode boundaries and the other electrode's currents have a much smaller contribution to the maximum current densities

\section{Conclusions}
If using optimal current patterns  one should  be sure to use the
right constraints. We suggest that in many situations the $L^\infty$
constraint may be the correct one. We have demonstrated that it is
simple to compute these optimal patterns, and the instrumentation
required to apply these patterns is much simpler than the $L^2$ or
power norm patterns. While still requiring multiple current sources,
they need only be able to switch between sinking and sourcing the
same current.

\begin{table}
 \begin{center}
  \begin{tabular}{|l|r|r|r|r|}
\hline
              & \multicolumn{2}{c|}{Single anomaly}& \multicolumn{2}{c|}{ Two anomalies} \\
\hline
Constraint & Voltage diff.& Power & Voltage diff.& Power     \\
\hline

$L^1$ Best pair drive           & 440.0911  & 1200.7618  &   347.3579   & 1185.4935   \\
$L^2$ optimal  	& 812.3243  & 3356.3035  &571.9161   &   3170.0985 \\
Power optimal & 518.9126  & 1768.3048  &321.616    &   1352.6896 \\
$L^\infty$ optimal & 1653.2673       & 19261.9208 &   968.2656   &   16798.3363 \\
\hline
  \end{tabular}
 \end{center}
\caption{\label{results}$L^2$ norm of voltage difference, and dissipated power for one and two anomalies with a variety
of constraints. The constraint levels have been chosen so that the
maximum electrode current is the same on each}
\end{table}

%The data for the one anomaly case is (cih = [0.51 0]; rih = 0.38;)
%The data for the two anomalies case is (cih=[0.7 0] and cih=[0 0.7], in both anomalies rih=0.2)

\begin{figure} 
 \begin{center}
  \includegraphics{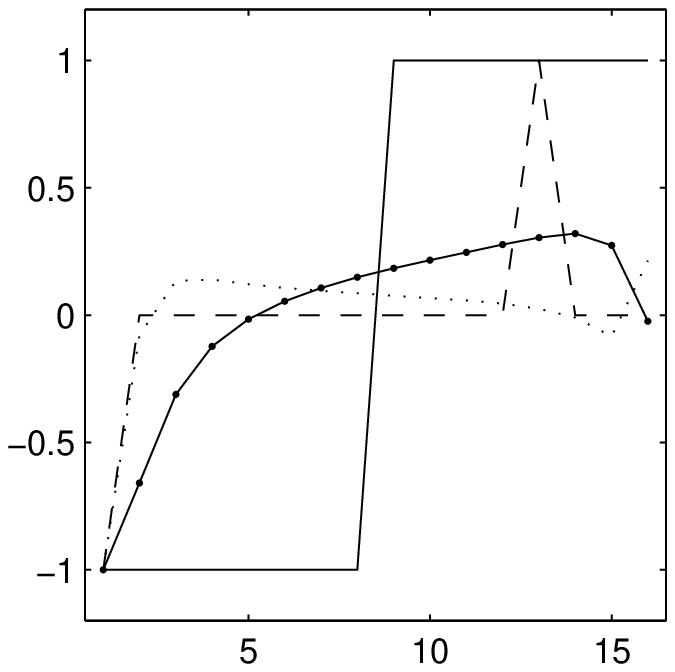}\includegraphics{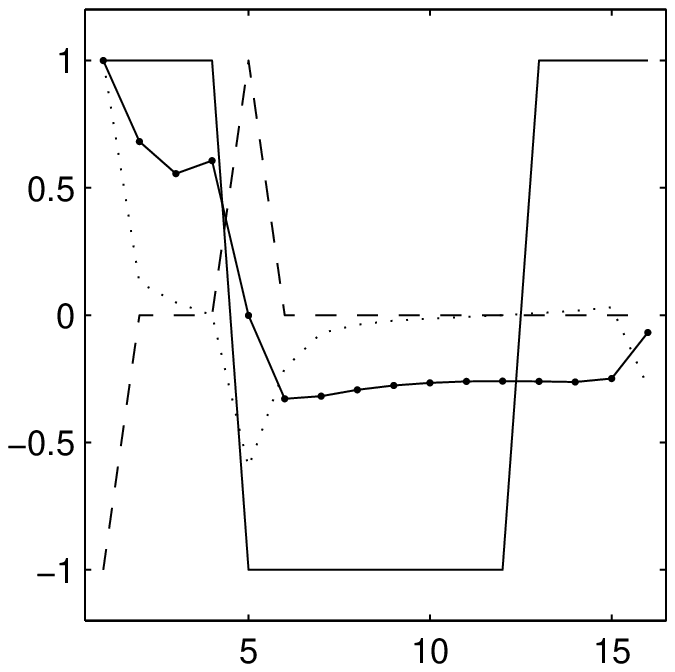}
 \end{center}
 \caption{\label{optcur}   
 Optimal current patterns.  Continuous line is the $L^\infty$
  norm, $-\circ-$ is the $L^2$ optimal, $\cdots$ power norm
  optimal and $--$ is  $L^1$ optimal (pair drive).}
\end{figure}
\begin{figure}   
 \begin{center}
   \includegraphics{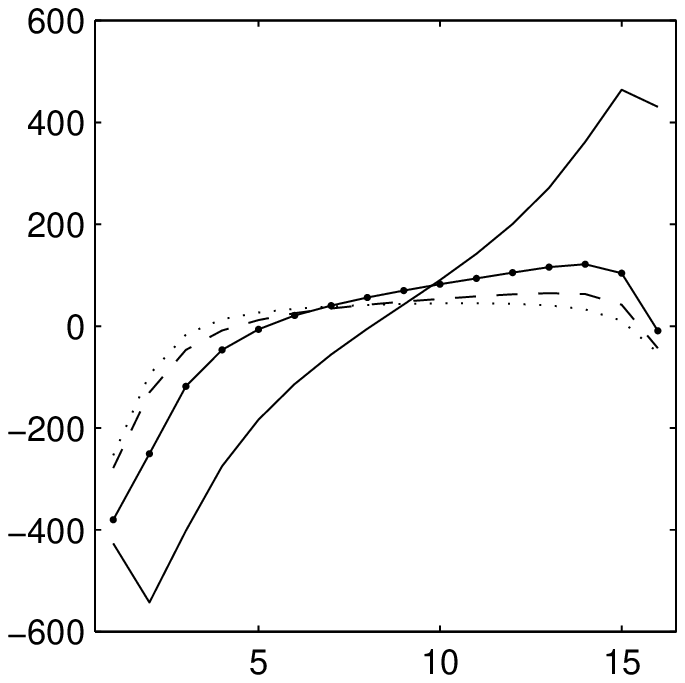}\includegraphics{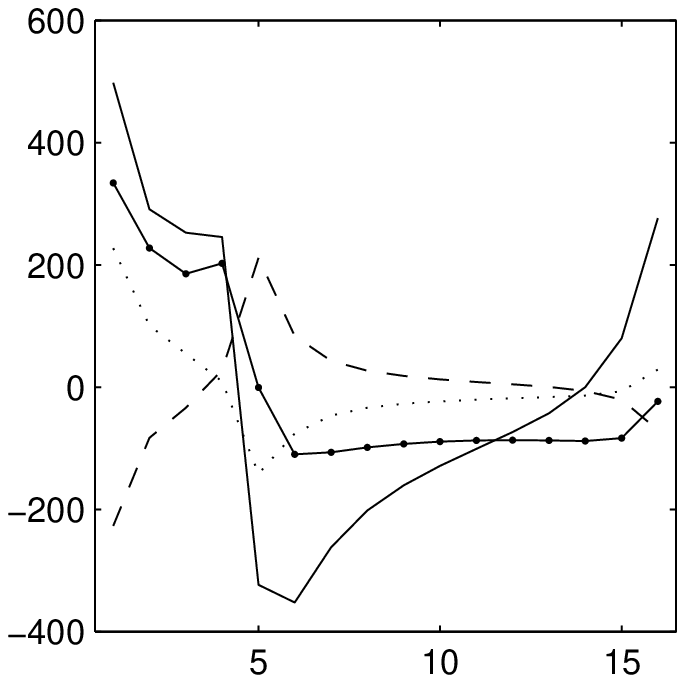}
 \end{center}
 \caption{\label{voltpat}Voltage difference measurements for one and two anomalies. For key see
   figure~\ref{optcur}}
\end{figure}
 
 %\bibliographystyle{unsrt}
%\bibliography{nirref/abbrsrc,nirref/ref,extra,local}

\end{document}